\documentclass[aps,pre,twocolumn,showpacs,floatfix]{revtex4-1}
\usepackage{graphicx}% Include figure files
\usepackage{dcolumn}% Align table columns on decimal point
\usepackage{bm}% bold math
\usepackage{epsf}
\usepackage{subfigure}
\usepackage{epstopdf}
\usepackage{amsmath}
\usepackage{amssymb}
\usepackage[cspex,bbgreekl]{mathbbol}
\usepackage{color}
\usepackage{stackrel}
\newcommand{\be}{\begin{equation}}
\newcommand{\ee}{\end{equation}}
\newcommand{\ba}{\begin{eqnarray}}
\newcommand{\ea}{\end{eqnarray}}
\newcommand{\s}{\boldsymbol \sigma}
\newcommand{\M}{{\bf M}}
\newcommand{\n}{{\bf n}}

\newcommand{\m}{{\bf m}}
\newcommand{\B}{{\bf B}}
\newcommand{\sr}{{\bf r}}
\graphicspath{{Code-Python/}}

\begin{document}

\title{Isometric fluctuation relations for equilibrium states \\
with broken symmetry}

\author{D. Lacoste$^1$}
\author{P. Gaspard$^2$}

\affiliation{$^1$ Laboratoire de Physico-Chimie Th\'eorique - UMR CNRS Gulliver 7083,\\ PSL Research University, ESPCI, 10 rue Vauquelin, F-75231 Paris, France\\
$^2$ Center for Nonlinear Phenomena and Complex Systems, \\Universit\'e Libre de Bruxelles, Code Postal 231, Campus Plaine, B-1050 Brussels, Belgium}

\date{\today}

\begin{abstract}
We derive a set of isometric fluctuation relations, which constrain the  
 order parameter fluctuations in finite-size systems at equilibrium and in the presence of a broken symmetry. 
These relations are exact and should apply generally to many condensed-matter physics systems. 
Here, we establish these relations for magnetic systems and nematic liquid crystals in a symmetry-breaking external field, and we illustrate them on the Curie-Weiss and the $XY$ models.  
Our relations also have implications for spontaneous symmetry breaking, which are discussed.
\end{abstract}


\maketitle
%%%%%%%%%%%%%%%%%%%%%%%%%%%%%%%%%%%%%%%%%%%%%%%%%%%%%%%%%%%%%%%%%%%%%

Away from equilibrium, the second law of thermodynamics quantifies the breaking of the 
time-reversal symmetry due to energy dissipation, as observed on macroscales. 
Yet, the microscopic equations of motion are fully reversible
and this microreversibility has fundamental implications such as the Onsager reciprocal relations
in regimes close to equilibrium, as well as the so-called fluctuation relations, 
which are also valid further away from equilibrium \cite{ECM93,GC95,K98,C99,AG06JSM,J11,S12}.

In this context, Hurtado {\it et al.} have uncovered a remarkable
extension, which they dubbed {\it isometric fluctuation relations}, by considering the symmetry under
both time reversal and spatial rotations in nonequilibrium fluids \cite{HPPG11}.
These results hint at the possibility that all the fundamental symmetries continue to 
manifest themselves in the fluctuations, even if these symmetries are broken by external constraints.
This concerns not only systems driven away from equilibrium, but also equilibrium systems
described by Gibbsian canonical distributions, 
as shown by one of us for discrete symmetries such as spin reversal \cite{G12}.
In this regard, we may wonder whether such fluctuation relations 
would hold for general broken symmetries in equilibrium systems.  
The issue is of importance given the central role played by symmetry-breaking phenomena
in physics \cite{A72,CL95}.  A symmetry may be broken spontaneously if the ground state has a lower symmetry than the Hamiltonian, 
or explicitly if a perturbation $H_1$ is added to a Hamiltonian $H_0$ where $H_1$ is less symmetric than $H_0$.  
In either case, do the fluctuations of the order parameter 
leave a footprint of the symmetry that is broken?

The purpose of the present paper is to answer this fundamental question in the affirmative by proving that, for equilibrium systems, 
whenever a symmetry is broken by an external field, the probability distribution of the fluctuations obey an isometric fluctuation relation.  
Remarkably, this relation is exact already for finite systems.
This result established for magnetic systems and nematic liquid crystals, is illustrated 
for the Curie-Weiss and $XY$ models of ferromagnetism.
We then discuss implications of our
result for spontaneous symmetry breaking (SSB).

Let us consider a system composed of $N$ Heisenberg spins $\s=\{ \pmb{\sigma}_i \}_{i=1}^N$, where the individual spins take discrete or continuous values such that $\pmb{\sigma}_i\in{\mathbb R}^d$ and 
$\Vert\pmb{\sigma}_i\Vert=1$.  The order parameter of this system is the magnetization 
$\M_N(\s)=\sum_{i=1}^N \s_i$, and the Hamiltonian of the system is assumed to be of the form $H_N(\s;\B)=H_N(\s;{\bf 0})- \B \cdot \M_N(\s)$ where $\B$ is the magnetic field.  Let us also introduce the probability $P_\B(\M)$ that the magnetization takes the value $\M=\M_N(\s)$ as $P_\B(\M) = \langle \delta\left[\M-\M_N(\s)\right]\rangle_\B$,
where $\delta(\cdot)$ denotes the Dirac delta distribution and $\langle \cdot\rangle_\B$ the statistical average over Gibbs' canonical measure at the inverse temperature~$\beta$ \cite{P03}.  
First, we establish a general identity between the distribution of the order parameter in the field, $P_{\bf B}({\bf M})$, and the same distribution in the absence of the field, $P_{\bf 0}({\bf M})$:
\ba
P_\B(\M) &=& \frac{1}{Z_N({\bf B})} \sum_{\s} {\rm e}^{-\beta H_N(\s;{\bf 0})+\beta \B \cdot \M_N(\s)} \delta\left[\M-\M_N(\s)\right], \nonumber\\
&=& \frac{1}{Z_N({\bf B})}\; {\rm e}^{\beta \B \cdot \M} \sum_{\s} {\rm e}^{-\beta H_N(\s;{\bf 0})} \delta\left[\M-\M_N(\s)\right],  
\nonumber\\
&=&\frac{Z_N({\bf 0})}{Z_N({\bf B})}\; {\rm e}^{\beta \B \cdot \M} \; P_{\bf 0}({\bf M}),
\label{GI}
\ea
where $Z_N(\B)$ is the partition function.  We notice that this identity holds even if the Hamiltonian $H_N(\s;{\bf 0})$ has no particular symmetry.

Now, we suppose that, in the absence of 
field, the Hamiltonian $H_N(\s;{\bf 0})$ is invariant under a symmetry group~$G$, which can be discrete or continuous.  This means that $H_N(\pmb{\sigma}^g;{\bf 0})= H_N(\pmb{\sigma};{\bf 0})$, where $\s^g=\{ {\boldsymbol{\mathsf R}}_g\cdot\pmb{\sigma}_i \}_{i=1}^N$, and ${\boldsymbol{\mathsf R}}_g$ is a representation of the member~$g$ of the group~$G$ such that $\vert\det{\boldsymbol{\mathsf R}}_g\vert=1$.  The consequence is that the probability distribution of the magnetization has this symmetry in the absence of magnetic field since summing over the microstates ${\s}$ or their symmetry transforms ${\s^g}$ are equivalent  for every $g\in G$ so that
\ba
P_{\bf 0}({\bf M})&=& \frac{1}{Z_N({\bf 0})} \sum_{\s^g} {\rm e}^{-\beta H_N(\s^g;{\bf 0})} \delta\left[\M-\M_N(\s^g)\right], \nonumber\\
&=& \frac{1}{Z_N({\bf 0})} \sum_{\s} {\rm e}^{-\beta H_N(\s;{\bf 0})} \delta\left[\M-{\boldsymbol{\mathsf R}}_g\cdot\M_N(\s)\right], \nonumber\\
&=& P_{\bf 0}({\boldsymbol{\mathsf R}}_g^{-1}\cdot{\bf M}).
\label{proof}
\ea
Combining Eqs.~(\ref{GI}) and~(\ref{proof}), 
one obtains the fluctuation relation:
\be
P_{\bf B}({\bf M}) = P_{\bf B}({\bf M'}) \ {\rm e}^{\beta \B\cdot({\bf M}-{\bf M'})}. 
\label{FT1}
\ee
with ${\bf M'}={\boldsymbol{\mathsf R}}_g^{-1}\cdot\M$ for all $g\in G$.  
When ${\boldsymbol{\mathsf R}}_g$ represents a rotation, $\Vert{\bf M}\Vert = \Vert{\bf M'}\Vert$, hence the name {\it isometric fluctuation relation}.  This relation includes as a particular case 
the fluctuation relation derived in Ref.~\cite{G12} when ${\bf M'}=-\M$. 
In analogy with the nonequilibrium case, a corollary of this relation can be obtained 
by introducing the Kullback-Leibler (KL) divergence of the distributions $P_{\bf B}(\M)$ and $P_{\bf B}({\bf M'})$. The positivity of the KL divergence leads to the second-law like inequality
\be
 \B \cdot \langle\M\rangle_\B \ge   \B \cdot \langle{\bf M'}\rangle_\B ,
\ee
where $\langle\cdot\rangle_\B$ represents an average with respect to the distribution $P_{\bf B}(\M)$.

In Eq.~(\ref{FT1}), the distribution of the order parameter is compared to the distribution 
of the rotated order parameter in the same magnetic field. Another possibility is to fix the 
order parameter and rotate the magnetic field. A similar derivation leads to:
\be
P_{\bf B}({\bf M}) = P_{\bf B'}({\bf M}) \ {\rm e}^{\beta ({\bf B}-{\bf B'})\cdot{\bf M}}, 
\label{FT2}
\ee
where ${\bf B'}={\boldsymbol{\mathsf R}}_g^{\rm T}\cdot\B$ for all $g\in G$, with $^{\rm T}$ denoting the transpose.  
We emphasize that the fluctuation relations~(\ref{FT1}) and~(\ref{FT2}) hold exactly in finite systems.

The isometric fluctuation relation~(\ref{FT1}) also holds locally for a spatially
varying magnetization density ${\bf m}(\sr)$ and magnetic field $\B(\sr)$. To show this, it is needed 
to proceed by coarse graining the magnetization density 
 ${\bf m}(\sr)=\sum_{i=1}^N \s_i\, \delta(\sr - \sr_i)$ where $\sr_i$ is the location 
of spin~$\s_i$. By adapting the derivation of Eq.~(\ref{FT1}), one then finds
\be
P_{\bf B}[{\bf m}(\sr)]=  P_{\bf B}[{\bf m'}(\sr)] \, {\rm e}^{\beta \int d\sr \, \B(\sr) 
\cdot [{\bf m}(\sr)-{\bf m'}(\sr) ]},
\label{FT-Mofr}
\ee  
where $P_{\bf B}[{\bf m}(\sr)]$ is the probability functional of the magnetization density ${\bf m}(\sr)$ and ${\bf m'}(\sr)={\boldsymbol{\mathsf R}}_g^{-1} \cdot{\bf m}(\sr)$ (see the Supplementary Material \cite{SM} for detail).

In the infinite-system limit $N\to\infty$, these fluctuation relations have their counterparts in terms of 
large-deviation functions \cite{E95,D07,T09}. By defining the magnetization per spin $\m=\M/N$, one can 
introduce a large-deviation function $\Phi_{\bf B}(\m)$ such that
\be
P_{\bf B}(\M) = A_N(\m)\, {\rm e}^{- N \Phi_{\bf B}(\m)} .
\label{LDP}
\ee
where $A_N(\m)$ is a prefactor which has a negligible contribution to $\Phi_{\bf B}(\m)$ 
in the limit $N \rightarrow \infty$.  As a result, Eq.~(\ref{FT1}) implies the following symmetry relation for the large-deviation function:
\be
\Phi_{\bf B}(\m) - \Phi_{\bf B}({\bf m'}) = \beta \B \cdot \left( {\bf m'} - \m \right).
\label{FT-LDP}
\ee
It is important to appreciate that the function $\Phi_{\bf B}(\m)$ characterizes 
the equilibrium fluctuations of the order parameter 
which are in general non Gaussian. This function can be expressed in terms of the Helmholtz free energy per spin,
$f(\B)=-\beta^{-1} \ln Z_N(\B)/N$, and of its Legendre transform $\Phi_{\bf 0}(\m)$
using Eqs.~(\ref{GI}) and~(\ref{LDP}), as
$\Phi_{\bf B}(\m)=\Phi_{\bf 0}(\m)-\beta\, \B \cdot \m - \beta  f(\B) + \beta f({\bf 0})$ \cite{G12}. 
Unlike the Helmholtz free energy $f(\B)$ or its Legendre transform $\Phi_{\bf 0}(\m)$, 
the function $\Phi_{\bf B}(\m)$ depends on both thermodynamically conjugated variables $\m$ and $\B$.

We may also introduce the cumulant generating function for the magnetization:
\be
\Gamma_{\bf B}(\pmb{\lambda}) \equiv \lim_{N\to\infty}-\frac{1}{N}\ln \left\langle{\rm e}^{-\pmb{\lambda}\cdot{\bf M}_N(\pmb{\sigma})}\right\rangle_{\bf B} ,
\label{GF}
\ee
which is the Legendre-Fenchel transform of the function $\Phi_{\bf B}(\m)$ defined by Eq.~(\ref{LDP}).
As a consequence of the isometric fluctuation relation~(\ref{FT1}), 
the generating function~(\ref{GF}) obeys the symmetry relation
$\Gamma_{\bf B}(\pmb{\lambda}) = \Gamma_{\bf B}\left[\beta{\bf B}+{\boldsymbol{\mathsf R}}_g^{\rm T}\cdot(\pmb{\lambda}-\beta{\bf B})\right]$ 
for all $g\in G$.  In the particular case of the inversion ${\boldsymbol{\mathsf R}}_g=-{\boldsymbol{\mathsf 1}}$, 
we find that
\be
\Gamma_{\bf B}(\pmb{\lambda}) = \Gamma_{\bf B}(2\beta{\bf B}-\pmb{\lambda}).
\label{FT-GF}
\ee
The first cumulant, which is the average magnetization per spin, is thus given by
\be
\langle{\bf m}\rangle_{\bf B} = \frac{\partial \Gamma_{\bf B}}{\partial\pmb{\lambda}}({\bf 0}) = -\frac{\partial \Gamma_{\bf B}}{\partial\pmb{\lambda}}(2\beta{\bf B}),
\label{FT-GF-M}
\ee
which has fundamental implications about SSB.
Indeed, as long as the cumulant generating function~(\ref{GF}) remains analytic in the variables $\pmb{\lambda}$
(which is necessarily the case in a finite system), the average magnetization has to vanish in the absence of external field
because $\langle{\bf m}\rangle_{\bf 0} = \partial_{\pmb{\lambda}}\Gamma_{\bf 0}({\bf 0}) = 
-\partial_{\pmb{\lambda}}\Gamma_{\bf 0}({\bf 0})=-\langle{\bf m}\rangle_{\bf 0}=0$, as implied by Eq.~(\ref{FT-GF-M}).
Due to the thermodynamic limit $N\to\infty$, the generating function may not be analytic, allowing
a spontaneous magnetization $\langle{\bf m}\rangle_{\bf 0}\neq 0$ in the absence of external field, and thus the possibility of SSB.
In this case, in view of the symmetry~(\ref{FT-GF}), 
the simplest possible form of the generating function near $\pmb{\lambda}=\beta{\bf B}$ is
\be
\big\vert \Gamma_{\bf B}(\pmb{\lambda})-\Gamma_{\bf B}(\beta{\bf B})\big\vert_{T_c} \sim \Vert\pmb{\lambda}-\beta{\bf B}\Vert^{1+1/\delta}
\label{USB}
\ee
at the critical temperature $T_c$, in order for the critical magnetization to scale as 
$\Vert\langle{\bf m}\rangle_{\bf B}\Vert_{T_c}\sim\Vert{\bf B}\Vert^{1/\delta}$ with the critical exponent $\delta=3$ in the 
mean-field models, or $\delta=15$ in the two-dimensional Ising model \cite{W65,F67,KGHHLPRSAK67}. 
The universal scaling behavior~(\ref{USB}) establishes the non-analyticity of the generating function, which allows for the 
possibility of a non-vanishing spontaneous magnetization in the thermodynamic limit.

\begin{figure}[h!]
%\begin{center}
\includegraphics[scale=0.3]{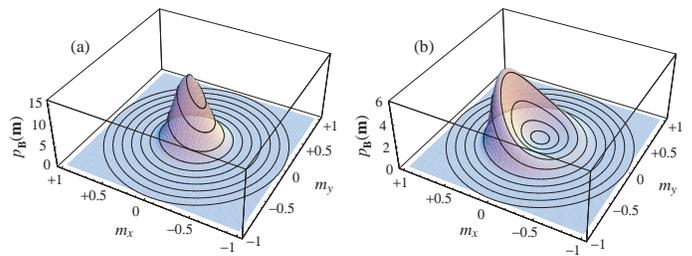}
\caption{Probability density $p_\B(\m)=N^3 P_\B(N\m)$ of the magnetization per spin $\m=\M/N=(m_x,m_y,m_z=0)$ for the three-dimensional Curie-Weiss model in the magnetic field $\B=(B,0,0)$ with $B=0.01$, $J=1$, and $N=100$ at the rescaled inverse temperatures (a) $\beta J=2.7$ in the paramagnetic phase and (b) $\beta J=3.3$ in the ferromagnetic phase.  The lines depict the contours of $\Vert\m\Vert=0.1,0.2,...,1.0$ where the isometric fluctuation relation (\ref{FT1}) holds \cite{SM}.}
\label{fig1}
%\end{center}
\end{figure}

\begin{figure}[h]
%\begin{center}
\includegraphics[scale=0.35]{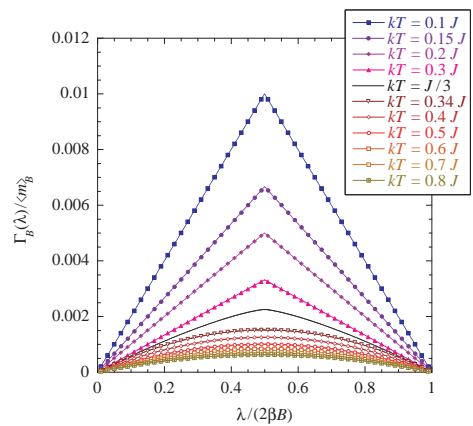}
\caption{Cumulant generating function (\ref{GF}) of the magnetization in the three-dimensional Curie-Weiss model for the magnetic field $\B=(B,0,0)$ with $B=0.001$, $J=1$, and different temperatures across criticality.  The generating function is rescaled by the average magnetization $\langle m\rangle_B$ in the direction of the external field and plotted versus the rescaled parameter $\lambda/(2\beta B)$ in the same direction $\pmb{\lambda}=(\lambda,0,0)$.  The generating function is computed by taking the Legendre-Fenchel transform of the large-deviation function $\Phi_{\bf B}({\bf m})$ introduced in Eq.~(\ref{LDP}) \cite{SM}.  The isometric fluctuation relation (\ref{FT1}) implies the symmetry $\lambda\to2\beta B-\lambda$ of the generating function according to Eq.~(\ref{FT-GF}).}
\label{fig2}
%\end{center}
\end{figure}

Now, we study a selection of illustrative examples of magnetic systems. 
Let us start by considering $N$ Heisenberg spins with Curie-Weiss interaction. The Hamiltonian of this system is
\be
H_N(\pmb{\sigma};{\bf B})=-\frac{J}{2N} \, {\bf M}_N(\pmb{\sigma})^2 -{\bf B}\cdot{\bf M}_N(\pmb{\sigma}).
\ee
The distribution of the order parameter is
\be
P_{\bf B}({\bf M})=\frac{1}{Z_N({\bf B})} \ {\rm e}^{\frac{\beta J}{2N}\, {\bf M}^2 +\beta{\bf B}\cdot{\bf M}}  C_N({\bf M}), 
\label{P(M)}
\ee
where the function $C_N(\M)=\sum_{\s} \delta\left[\M-\M_N(\s)\right]$
represents the number of microstates with a given magnetization $\M$. This number, which has to be rotationally invariant, is related by $C_N={\rm e}^{S_N/k}$ to the entropy function $S_N(\M)$ and Boltzmann constant $k$.
Using large-deviation theory \cite{SM,E95,T09}, one explicitly obtains this entropy in the form of $S_N(N\m)=-k N I(m)$ with $m=\Vert\m\Vert$ and
\be
I(m)=m \mathcal{L}^{-1} (m)- \ln \frac{\sinh\left[\mathcal{L}^{-1}(m)\right]}{\mathcal{L}^{-1}(m)},
\label{I(m)}
\ee
where $\mathcal{L}^{-1}$ is the inverse of the Langevin function $\mathcal{L}(x)=\coth(x)-1/x$, 
a result which also follows from a standard mean-field approach \cite{LL01}. 
Combining Eqs.~(\ref{P(M)})-(\ref{I(m)}), the large-deviation function defined in Eq.~(\ref{LDP}) is: 
$\Phi_\B(\m)=I(m) - \beta J m^2/2 - \beta \B \cdot \m$. The prefactor $A_N(\m)$ of Eq.~(\ref{LDP}) is calculated in the Supplementary Material \cite{SM}.  For this model, it is straightforward to check that 
this large-deviation function satisfies the symmetry relation of Eq.~(\ref{FT-LDP}).
In Fig.~\ref{fig1}, we show the distribution $P_{\bf B}(N{\bf m})$ as a function of the  components $(m_x,m_y)$ of the magnetization per spin. 
In the absence of external field, the probability
distribution is spherically symmetric, in which case spontaneous symmetry breaking occurs below the critical temperature.  Figure~\ref{fig2} depicts the cumulant generating function~(\ref{GF}) below and above the critical temperature.  If this function is analytic in the paramagnetic phase above the critical temperature, it is no longer the case below the critical temperature in the ferromagnetic phase where the function presents a discontinuity at the symmetry point $\pmb{\lambda}=\beta{\bf B}$ in its derivatives with respect to the parameters $\pmb{\lambda}$.  As aforementioned, this non-analyticity is at the origin of the spontaneous magnetization in the ferromagnetic phase.  At the critical temperature, the generating function has the universal scaling behavior~(\ref{USB}) with $\delta=3$, as it should for this mean-field model \cite{SM}.

We proceed by investigating the more complex $XY$~model, in light of our
findings on isometric fluctuation relations.  In this much studied model, topological defects unbind above the Kosterlitz-Thouless transition temperature $T_{\rm KT}$, where
the order changes from quasi-long range to short range \cite{KT73}. The Hamiltonian of the $XY$~model in an external magnetic field ${\bf B}=(B_x,B_y)$ is given by
\be
H_N(\pmb{\theta};{\bf B})= - J\, \sum_{\langle i,j\rangle} \cos(\theta_i-\theta_j) - {\bf B}\cdot{\bf M} \, ,
\label{H-XY}
\ee
on a square lattice with $L\times L$ sites ($i,j=1,2,...,L$) with the magnetization ${\bf M} = \sum_i \left(\cos\theta_i,\sin\theta_i\right)$.  In the absence of external field, the Hamiltonian is symmetric under the orthogonal group O(2).  In view of Eq.~(\ref{GI}), the probability distribution of the magnetization in the field can be obtained from the same distribution in the absence of the field $P_{\bf 0}({\bf M})$. This quantity is itself related to the probability distribution of the modulus of the magnetization $Q(M)$ by  $P_{\bf 0}({\bf M})=Q(M)/(2 \pi M)$. A rich physics is contained in the distribution $Q(M)$ \cite{BHP98}.  In particular, a calculation of this quantity below $T_{\rm KT}$ has been shown to be numerically very close to a Gumbel distribution \cite{PHSB01}. 

 \begin{figure}[h]
%\begin{center}
\includegraphics[scale=0.25]{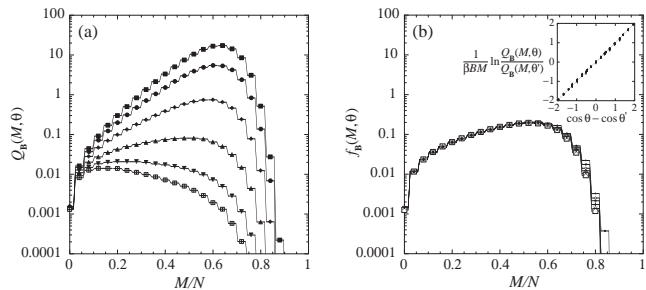}
\caption{$XY$~model of two-dimensional magnetism on a square lattice with $L=10$ and $J=1$ in the external magnetic field $\B=(0.1,0)$ at the rescaled 
inverse temperature $\beta J=0.8$ above the critical 
temperature $T_{\rm KT}$. (a) The distribution $Q_{\bf B}(M,\theta)\equiv 2 \pi M P_{\bf B}({\bf M})$ versus the modulus of the magnetization per spin $M/N=\Vert\M\Vert/N$ for different values of the angle $\theta$ separated by $\Delta\theta=\pi/6$, with the top curve corresponding to the direction along the magnetic field. (b) The distribution compensated by Boltzmann weights $f_{\bf B}(M,\theta)\equiv Q_{\bf B}(M,\theta) \exp(-\beta B M \cos\theta)$ confirming the isometric fluctuation relation.  Inset: Check of the equality between the left- and right-hand sides of Eq.~(\ref{log-FR}). After a transitory run of $10^4$ spin flips, the statistics is carried out over $10^8$ values of the magnetization, each one separated by $10^3$ spin flips.}
\label{fig3}
%\end{center}
\end{figure}

In order to test the isometric fluctuation relations, we have carried out Monte Carlo simulations as shown in Fig.~\ref{fig3}.  Figure~\ref{fig3}a represents the quantity $Q_{\bf B}(M,\theta)\equiv 2 \pi M P_{\bf B}({\bf M})$ versus the magnetization per spin $m=M/N$ for different values of the angle $\theta$, while Fig.~\ref{fig3}b depicts the probability distribution compensated by Boltzmann weights: $f_{\bf B}(M,\theta)\equiv Q_{\bf B}(M,\theta) {\rm e}^{-\beta B M \cos\theta}$.  The coincidence of all the curves is the evidence of the isometric fluctuation relation. In addition, we show in the inset a test of the relation using an equivalent form put forward by Hurtado {\it et al.} \cite{HPPG11} 
\be
\frac{1}{\beta B M}\, \ln\frac{Q_{\bf B}(M,\theta)}{Q_{\bf B}(M,\theta')} = \cos\theta-\cos\theta' \, .
\label{log-FR}
\ee
We have checked that the relation holds at temperatures below $T_{\rm KT}$, as well as above. 
 
Besides the Curie-Weiss and $XY$ models, the isometric fluctuation relation applies as well to other magnetic systems. In particular, the relation can be established using the transfer-matrix method in the case of a 1D classical chain of Heisenberg spins \cite{LG14}.
 
Beyond magnetic systems, broken symmetry phases are ubiquitous in soft matter systems, such as liquid crystals. These systems are of great interest to study deformations and orientation due to heterogeneities or to the application of external fields. Below, we focus on nematic liquid crystals which can be described by a tensorial order parameter ${\boldsymbol{\mathsf Q}}$, or equivalently by a scalar 
order parameter and a director $\n$ for uniaxial nematics \cite{dGP93}.  Here, we discuss the fluctuations of the tensorial order parameter in a finite ensemble of nematogens.  

Let us consider the following general Hamiltonian:
\be
H_N(\pmb{\sigma};{\bf B}) = H_N(\pmb{\sigma};{\bf 0}) - {\bf B}^{\rm T}\cdot {\boldsymbol{\mathsf Q}}_N (\pmb{\sigma}) \cdot {\bf B},
\label{H-Q}
\ee
with the following traceless tensorial order parameter 
\be
{\boldsymbol{\mathsf Q}}_N(\pmb{\sigma})=\sum_{i=1}^N \left(\pmb{\sigma}_i\otimes \pmb{\sigma}_i^{\rm T}-\frac{1}{d} \, {\boldsymbol{\mathsf 1}}\right)
\ee
where now $\pmb{\sigma}_i\in{\mathbb R}^d$ is a unit vector directed along the axis of the nematogens molecules.  The distribution of this tensor is
$P_{\bf B}({\boldsymbol{\mathsf Q}}) \equiv \langle \delta\left[{\boldsymbol{\mathsf Q}}-{\boldsymbol{\mathsf Q}}_N(\pmb{\sigma})\right]\rangle_{\bf B}$ where $\langle\cdot\rangle_{\bf B}$ denotes the statistical average over Gibbs' canonical measure.
Using a similar derivation as before for a vectorial order parameter, one obtains the following isometric fluctuation relations for the distribution of the tensorial order parameter ${\boldsymbol{\mathsf Q}}$:
\be
P_{\bf B}({\boldsymbol{\mathsf Q}}) = P_{\bf B}({\boldsymbol{\mathsf Q'}}) \ {\rm e}^{\beta\, {\bf B}^{\rm T}\cdot({\boldsymbol{\mathsf Q}}-{\boldsymbol{\mathsf Q'}})\cdot{\bf B}}
\label{FT-Q-1}
\ee
with ${\boldsymbol{\mathsf Q'}}={\boldsymbol{\mathsf R}}_g^{-1}\cdot {\boldsymbol{\mathsf Q}}\cdot{\boldsymbol{\mathsf R}}_g^{-1{\rm T}}$ for all $g\in G$.  
We will report elsewhere the application of this relation to a variant of the Maier-Saupe model \cite{MS58} 
and its extension to the continuum description of long-wavelength fluctuations of the director field $\n(\sr)$ \cite{LG14}.

In this paper, we have obtained isometric fluctuation relations for equilibrium systems.  These relations are exact and hold for 
finite as well as infinite systems.  We have shown that the fundamental symmetries of systems undergoing explicit or spontaneous symmetry breaking 
continue remarkably to manifest themselves in the fluctuations of the order parameter.
The fluctuation relation take slightly different forms depending on the particular interaction energy of the system with 
the symmetry breaking field, as shown in the examples with magnetic or nematic systems.

We have also shown in Eq.~(\ref{FT-Mofr}) that the symmetry relation holds not only 
for the global order parameter in a finite system but also locally in spatially extended systems. 
A potential application of this result for experiments could consist in looking for an asymmetry 
in the local fluctuations of an order parameter, 
and in extracting from this asymmetry, information about the symmetry breaking field $\B(\sr)$.
This would be the equilibrium analog of recent experiments, in which  
the asymmetry in nonequilibrium fluctuations have been exploited to estimate the 
entropy production \cite{PG2008,DL2014}.  
Another interesting application of this framework, which we plan to pursue, concern 
the analysis of fluctuations in the critical regime, which are accessible experimentally \cite{JPCG08}.
In conclusion, the isometric fluctuation relations
point towards a deep connection between fluctuations and symmetries, beyond the distinction between 
equilibrium and non-equilibrium. This deep link, not only brings a new light
on the classic topic of symmetry breaking, but is also likely to be a useful tool for extracting
relevant information from fluctuations.

%%%%%%%%%%%%%%%%%%%%%%%%%%%%%%%%%%%%%%%%%%%%%%%%%%%%%%%%%%%%%%%%%%%%%%
\section*{Acknowledgments}
The authors thank P.~Reimann for helpful advices in the presentation of this paper, as well as M.~Clusel, M.~Esposito, and P.~Davidson for stimulating discussions. P. Gaspard thanks the Belgian Federal Government for financial support under the Interuniversity Attraction Pole project P7/18~``DYGEST".

\vskip 10pt
%%%%%%%%%%%%%%%%%%%%%%%%%%%%%%%%%%%%%%%%%%%%

%%%%%%%%%%%%%%%%%%%%%%%%%%%%%%%%%%%%%%%%%%%%%%%%%%%%%%%%%%%%%%%%%%%%%
\begin{widetext}

\vskip 0.8 cm

\appendix{\centerline{\large\sf Supplementary Material}}

\vskip 0.3 cm

Here, we prove isometric fluctuation relations for the probability functional of the local order parameter or its Fourier components and we use large-deviation theory \cite{E95,T09} to evaluate the probability distributions of the order parameter in the three-dimensional Curie-Weiss model of ferromagnetism.

\section{The isometric fluctuation relation for the magnetization density}

The volume $V$ of the magnet is partitioned into small cells $\{\Delta V_j\}_{j=1}^c$ where the magnetization density ${\bf m}(\sr)=\sum_{i=1}^N\s_i\,\delta(\sr-\sr_i)$ is coarse grained as
\be
{\bf m}_j=\frac{1}{\Delta V_j}\int_{\Delta V_j} d\sr \, \m(\sr) .
\ee
Moreover, the external magnetic field $\B(\sr)$ is supposed to be piecewise constant in the cells: $\B(\sr)=\B_j$ for $\sr\in\Delta V_j$.  The joint probability distribution of the magnetization per spin in the cells is thus introduced as
\be
P_\B\left(\{\m_j\}\right) \equiv \left\langle\prod_{j=1}^{c} \delta\left[ \m_j-\frac{1}{\Delta V_j}\int_{\Delta V_j} d\sr \, \m(\sr)\right]\right\rangle_\B \, ,
\ee
where $\langle\cdot\rangle_\B$ denotes the statistical average over Gibbs' canonical probability distribution of Hamiltonian $H=H_0+H_{\rm ext}$ where $H_0$ is invariant under rotations. The interaction with the external field can be written as
\be
H_{\rm ext} = - \int_V d\sr \, \B(\sr) \cdot \m(\sr) = -\sum_{j=1}^c \int_{\Delta V_j} d\sr \, \B(\sr) \cdot \m(\sr) = -\sum_{j=1}^c \B_j \cdot \m_j \, \Delta V_j \, ,
\ee
so that the joint probability distribution takes the following form:
\be
P_\B\left(\{\m_j\}\right) = \frac{Z_N({\bf 0})}{Z_N({\bf B})}\; {\rm e}^{\beta\sum_{j=1}^c \B_j \cdot \m_j \, \Delta V_j} \; P_{\bf 0}\left(\{\m_j\}\right) .
\ee
Since the Hamiltonian $H_0$ is symmetric under the group $G$ of rotations, we obtain the isometric fluctuation relation
\be
P_\B\left(\{\m_j\}\right) = P_\B\left(\{{\bf m'}_j\}\right) \, {\rm e}^{\beta\sum_{j=1}^c \B_j \cdot (\m_j-{\bf m'}_j) \, \Delta V_j} \, ,
\ee
where ${\bf m'}_j={\boldsymbol{\mathsf R}}_g^{-1} \cdot{\bf m}_j$ for $g\in G$, and $\Vert{\bf m'}_j\Vert=\Vert{\bf m}_j\Vert$.

In the limit where the cells of the partition are arbitrarily small, the joint probability distribution becomes the probability functional of the magnetization density and the sum becomes an integral so that the isometric fluctuation relation reads
\be
P_\B[\m(\sr)] = P_\B[{\bf m'}(\sr)] \, {\rm e}^{\beta\int_V d\sr \B(\sr) \cdot [\m(\sr)-{\bf m'}(\sr)]} \, ,
\ee
with ${\bf m'}(\sr)={\boldsymbol{\mathsf R}}_g^{-1} \cdot{\bf m}(\sr)$ for $g\in G$, as announced.

\section{The Curie-Weiss model of ferromagnetism}

In this three-dimensional model, the interaction between $N$ Heisenberg spins $\pmb{\sigma}_i=(\sin\theta_i\cos\phi_i,\sin\theta_i\sin\phi_i,\cos\theta_i)$ is ruled by the Hamiltonian
 \be
H_N(\pmb{\sigma};{\bf B})=-\frac{J}{2N} \, {\bf M}_N(\pmb{\sigma})^2 -{\bf B}\cdot{\bf M}_N(\pmb{\sigma})
\ee
where
\be
{\bf M}_N(\pmb{\sigma})=\sum_{i=1}^N \pmb{\sigma}_i 
\label{magnet}
\ee
is the total magnetization, which is {\it a priori} distributed according to
\be
C_N({\bf M}) \equiv \int \frac{d^N\pmb{\sigma}}{(4\pi)^N} \ \delta\left[{\bf M}-{\bf M}_N(\pmb{\sigma})\right] 
\label{C-Heisenberg}
\ee
with $d^N\pmb{\sigma}=\prod_{i=1}^N d\cos\theta_i d\phi_i$.  This distribution is normalized according to
\be
\int C_N({\bf M})\, d{\bf M}=1
\ee
where $d{\bf M}=M^2dM d\cos\theta d\phi$ in spherical coordinates.

We apply large-deviation theory \cite{E95,T09} in order to obtain the behavior of this distribution as
\be
C_N(N{\bf m}) = A_N(m) \, {\rm e}^{-N I(m)} \qquad\mbox{for}\quad N\to\infty
\label{Hyp1}
\ee
in terms of some rate function $I(m)$ and a sub-exponential prefactor $A_N(m)$.  Because of 
the rotational invariance of $C_N(N{\bf m})$, the rate function and the prefactor only depend 
on the modulus of the magnetization per spin $m=\Vert {\bf m}\Vert=\Vert {\bf M}\Vert/N$.  
For the purpose of deducing $I(m)$ and $A_N(m)$, we introduce the generating function of the statistical moments of the magnetization:
\be
\tilde C_N({\bf h}) \equiv \left\langle {\rm e}^{{\bf h}\cdot{\bf M}_N(\pmb{\sigma})}\right\rangle = \int d{\bf M}\, {\rm e}^{{\bf h}\cdot{\bf M}} \, C_N({\bf M}) .
\label{C(h)-asympt}
\ee
Since the magnetization is defined as the sum~(\ref{magnet}) over spins that are statistically independent according to the distribution~(\ref{C-Heisenberg}), this generating function is given by
\be
\tilde C_N({\bf h})= \chi(h)^N
\label{C-chi-N}
\ee
with
\be
\chi(h) = \int \frac{d\pmb{\sigma}}{4\pi} \, {\rm e}^{{\bf h}\cdot\pmb{\sigma}} =\frac{\sinh h}{h} .
\ee
Now, Eq.~(\ref{Hyp1}) is inserted into Eq.~(\ref{C(h)-asympt}) and the integral over ${\bf M}=N{\bf m}$ is carried out in spherical coordinates with the method of steepest descent~\cite{BH86}.
In this way, the generating function is obtained as
\be
\tilde C_N({\bf h}) \simeq A_N(m_h) \, \frac{(2\pi N)^{3/2} m_h}{h\sqrt{I''(m_h)}} \, {\rm e}^{N[h m_h - I(m_h)]} \qquad\mbox{for}\quad N\to\infty
\label{C(h)-formula}
\ee
in terms of the root $m_h$ of
\be
h = \frac{dI}{dm}(m_h) \, .
\ee
Equating (\ref{C-chi-N}) to (\ref{C(h)-formula}), the rate function is thus determined as
\be
I(m) = m h_m - \ln \frac{\sinh h_m}{h_m} 
\ee
where $h_m$ is the root of
\be
m=\frac{d}{dh}\ln\chi(h_m) = {\cal L}(h_m)
\ee
in terms of the Langevin function ${\cal L}(h)=\coth(h)-1/h$.  Inversely, we have that
\be
h_m = {\cal L}^{-1}(m) = I'(m) ,
\ee
hence 
\be
I(m)=m \mathcal{L}^{-1} (m)- \ln \frac{\sinh\left[\mathcal{L}^{-1}(m)\right]}{\mathcal{L}^{-1}(m)} .
\label{I(m)-SM}
\ee

We notice that the rate function is equivalently given by the Legendre-Fenchel transform 
\be
I(m) = {\rm Max}_h\left[ mh-\ln\chi(h)\right] ,
\label{rate fct}
\ee
according to the G\"artner-Ellis theorem \cite{E95,T09}.  
Interestingly, there is another route to arrive at this result, which amounts to optimize the 
free energy function $mh-\ln\chi(h)$ with respect to an unknown probability density of the order parameter $\rho(\m,\Omega)$ in the solid
angle $\Omega$, rather than with respect to $h$. This is the essence of the variational mean-field approach \cite{CL95}.  For the present case, we can implement this method by introducing the average of the magnetization $\m$ by
\be
\langle \m \rangle = \int d \Omega \, \m \, \rho(\m,\Omega),
\label{avera m}
\ee
where $\rho(\m,\Omega)$ is a probability distribution, which
also depends on the applied magnetic field and is yet to be determined.
Now, the quantity $C_N({\bf M})$ introduced above, has the interpretation of the number of spin configurations 
with the given magnetization $\M$. Therefore it can only depend on $M=\Vert\M\Vert$ and is related to the
rotational Shannon entropy $S_N(N \m)$ of the single particle distribution $\rho(\m,\Omega)$,
 by Boltzmann formula: $C_N=e^{S_N/k}$, where
\be
S_N(N\m)= -N k \int d \Omega\, \rho(\m,\Omega)\, \ln \rho(\m,\Omega).
\label{rot entropy}
\ee
It follows from this that a suitable mean-field free energy can be written as 
$F_{\rm MF}[\rho(\m,\Omega)]=E_N(N\langle{\bf m}\rangle;{\bf B})- T S_N(N \m)$ where the mean energy $E_N/N=-(J/2)\langle{\bf m}\rangle^2-{\bf B}\cdot\langle{\bf m}\rangle$ is expressed in terms of the average magnetization~(\ref{avera m}).  By its extensivity, this free energy is of the form $N f_{\rm MF}[\rho(\m,\Omega)]$.  From the equation obtained by imposing that the functional derivative of $f_{\rm MF}[\rho(\m,\Omega)]$ with respect to $\rho(\m,\Omega)$ be zero, one obtains the optimal single particle distribution solution of the variational problem, $\rho_{\rm MF}(\m,\Omega)$. The
solution has the form $\rho_{\rm MF}(\m,\Omega) \sim \exp({\bf h} \cdot \m)$ where ${\bf h}$ is the 
mean field ${\bf h}=\beta J\langle \m \rangle + \beta\B$. This vector is directed along $z$, so that we can use $h={\bf h} \cdot {\bf e}_z$ and $m=\m \cdot {\bf e}_z$. Then,
one obtains from Eq. (\ref{avera m}) a self-consistent equation, which
is $m= {\cal L}(h)$. Thus, in the notation introduced above $h=h_m$.
It is then a simple matter, to insert this result into Eq. (\ref{rot entropy}), and to prove that
 $S_N(N \m)=-N k I(m)$ where $I(m)$ is the rate function of Eq. (\ref{I(m)-SM}). Thus, the rate function 
 represents the rotational entropy, which is also the Shannon entropy 
 of the one-particle distribution, while the large-deviation function represents the 
 free energy per spin for a given value of $\m$ divided by $\beta^{-1}=kT$.

While the variational mean-field or the G\"artner-Ellis theorem leads to the same
rate function as also obtained by the explicit calculation via Eqs. (\ref{magnet})-(\ref{I(m)-SM}), 
one advantage of the latter method is that one obtains the prefactor of the large-deviation function
which the other methods do not give. For the present example, one obtains:
\be
A_N(m) \simeq \frac{h_m\sqrt{I''(m)}}{(2\pi N)^{3/2}\, m} 
\ee
with
\be
I''(m) = \frac{1}{{\cal L}'(h_m)} = \left(\frac{1}{h_m^2}-\frac{1}{\sinh^2 h_m}\right)^{-1} .
\ee

Finally, the probability distribution function is obtained as
\be
C_N(Nm)\simeq \frac{{\cal L}^{-1}(m)}{(2\pi N)^{3/2}\,m}\left[\frac{1}{{\cal L}^{-1}(m)^2}-\frac{1}{\sinh^2 {\cal L}^{-1}(m)}\right]^{-1/2} \, {\rm e}^{-NI(m)} \qquad\mbox{for}\quad N\to\infty
\label{C(m)-asympt}
\ee
with the rate function~(\ref{I(m)-SM}).  The latter can be expanded in power series as
\be
I(m) = \frac{3}{2}\, m^2 +\frac{9}{20}\, m^4 +\frac{99}{350}\, m^6 +\frac{1539}{7000}\, m^8 +\frac{126117}{673750}\, m^{10} + O(m^{12}) \, ,
\ee
which diverges at $m=1$ in accordance with the fact that the magnetization is bounded as $\Vert{\bf M}\Vert = Nm \leq N$.  We also have that
\be
C_N(0) = \left(\frac{3}{2\pi N}\right)^{3/2} \, ,
\ee
since $I''(0)=3$ and $\lim_{m\to 0} h_m/m = \lim_{m\to 0} I'(m)/m=3$.

At finite temperature and in the presence of an external magnetic field, the probability distribution of the magnetization is thus given by
\be
P_{\bf B}({\bf M}) = \frac{1}{Z_N(B)} \ {\rm e}^{N\beta\left(\frac{J}{2}\, m^2 +Bm\cos\theta\right)} \ C_N(Nm)
\label{P(m)-asympt}
\ee
where $\theta$ is the angle between $\bf B$ and ${\bf m}={\bf M}/N$.

For ${\bf B}=0$, the paramagnetic-ferromagnetic phase transition occurs at the critical value $\beta_{\rm c}J=3$, beyond which the probability distribution has its maximum at a non-zero value of the magnetization.

Equations (\ref{C(m)-asympt}) and (\ref{P(m)-asympt}) are used to plot Figs.~1a-1b of the main text.  These figures depict the probability density $p_{\bf B}({\bf m})=N^3 P_{\bf B}(N{\bf m})$ of the magnetization per spin. If ${\bf B}=(B,0,0)$, this density obeys the isometric fluctuation relation
\be
p_{\bf B}(m\cos\theta,m\sin\theta,0)=p_{\bf B}(m,0,0)\, {\rm e}^{N\beta B m (\cos\theta-1)}
\ee
along the lines at constant values of $m=\sqrt{m_x^2+m_y^2}$ in Figs.~1a-1b of the main text.

\begin{figure}[h]
%\begin{center}
\includegraphics[scale=0.5]{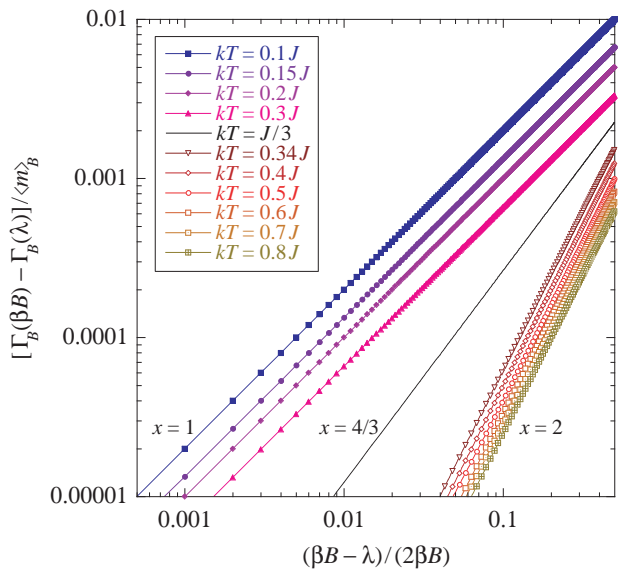}
\caption{Generating function of the magnetization cumulants in the three-dimensional Curie-Weiss model for the magnetic field $\B=(B,0,0)$ with $B=0.001$, $J=1$, and different temperatures across criticality.  The plot depicts the scaling behavior of the generating function close to its maximum at the symmetry point $\lambda=\beta B$.  The difference $\Gamma_B(\beta B)-\Gamma_B(\lambda)$ is rescaled by the average magnetization $\langle m\rangle_B$ and $\beta B-\lambda$ by $2\beta B$.}
\label{fig1-SM}
%\end{center}
\end{figure}

The large-deviation function is defined as
\be
\Phi_{\bf B}({\bf m})\equiv \lim_{N\to\infty}-\frac{1}{N}\ln P_{\bf B}(N{\bf m})
\label{LDF}
\ee
and its Legendre-Fenchel transform gives the cumulant generating function
\be
\Gamma_{\bf B}(\pmb{\lambda}) = {\rm Min}_{\bf m}\left[ \Phi_{\bf B}({\bf m})+\pmb{\lambda}\cdot{\bf m}\right].
\label{GF2}
\ee
Since $\Phi_{\bf B}(\m)=\Phi_{\bf 0}(\m)-\beta\, \B \cdot \m - \beta  f(\B) + \beta f({\bf 0})$ and $\Phi_{\bf 0}(\m)$ can be related to $f(\B)$ by
a Legendre-Fenchel transform, we find that $\Gamma_{\bf B}(\pmb{\lambda}) =\beta\left[ f(\B-\beta^{-1}\pmb{\lambda})-f(\B)\right]$.
Taking ${\bf B}=(B,0,0)$, ${\bf m}=(m,0,0)$, and $\pmb{\lambda}=(\lambda,0,0)$, the cumulant generating function presents a maximum at the symmetry point $\lambda=\beta B$, as shown in Fig.~2 of the main text.  The scaling behavior around this point is evidenced by plotting $\Gamma_B(\beta B)-\Gamma_B(\lambda)$ versus $\beta B-\lambda$, as shown in Fig.~\ref{fig1-SM}.  In this log-log plot, the slope provides the scaling exponent $x$ in the relation
\be
\Gamma_B(\beta B)-\Gamma_B(\lambda) \sim (\beta B-\lambda)^x .
\ee
In the paramagnetic phase for $kT>kT_c=J/3$, the generating function is analytic for $\lambda\in[0,2\beta B]$ and it presents a smooth quadratic maximum at $\lambda=\beta B$ where the scaling exponent is $x=2$.  Because of the analyticity, the average magnetization per spin is vanishing in the absence of external field: $\langle m\rangle_0=0$.
However, in the ferromagnetic phase for $kT<kT_c=J/3$, the generating function has a discontinuity in its derivative at its maximum so that the scaling exponent is now $x=1$.  This non-analyticity allows for a spontaneous magnetization $\langle m\rangle_0\neq 0$ in the ferromagnetic phase.  At criticality, the scaling behavior is intermediate with an exponent $x=4/3$ in this mean-field model.  This value confirms the conjecture of the main text that this critical exponent should be equal to $x=(\delta+1)/\delta$ where $\delta$ is the exponent between the average magnetization and the external field at the critical temperature $T_c$: $\langle m\rangle_{B,T_c}\sim B^{1/\delta}$ \cite{H87}.  Indeed, this exponent takes the value $\delta=3$ in mean-field models, so that $x=4/3$ as here observed.

%%%%%%%%%%%%%%%%%%%%%%%%%%%%%%%%%%%%%%%%%%%%%%%%%%%%%%%%%%%%%%%%%%%%%%
\section*{Acknowledgments}
The authors thank D. Gaspard for helpful advices in power expansions with Mathematica \cite{W88}.

\vskip 10pt
%%%%%%%%%%%%%%%%%%%%%%%%%%%%%%%%%%%%%%%%%%%%
\end{widetext}

\end{document}